# Measurements of Broadband Negative Index in Space-Coiling Acoustic Metamaterials

Yangbo Xie, Bogdan-Ioan Popa, Lucian Zigoneanu, and Steven A. Cummer*

*Department of Electrical and Computer Engineering, Duke University, Durham, North Carolina 27708, USA.*

We report the experimental demonstration of broadband negative refractive index obtained in a labyrinthine acoustic metamaterial structure. Two different approaches were employed to prove the metamaterial negative index nature: one-dimensional extractions of effective parameters from reflection and transmission measurements, and two-dimensional prism-based measurements that convincingly show the transmission angle corresponding to negative refraction. The transmission angles observed in the latter case also agree very well with the refractive index obtained in the one-dimensional measurements and numerical simulations. We expect this labyrinthine metamaterial to become the unit cell of choice for practical acoustic metamaterial devices that require broadband and significantly negative indexes of refraction.

Acoustic metamaterials are artificial materials designed to exhibit unusual acoustic material parameters, such as negative refractive index [1] or strong effective mass anisotropy [2]. Acoustic metamaterials not only offer the ability to manipulate sound and vibrations in unprecedented ways, but also offer promising applications (especially when combined with the concept of transformation acoustics [3-5]), such as acoustic cloaking[6] and other transformation devices [4], subwavelength imaging [7-9], transmission/reflection control [10], and surface wave manipulation[11]. Creating negative acoustic parameters is one of the primary goals of acoustic metamaterial research. This has been demonstrated with Helmholtz resonator based metamaterials [12], membrane-based materials [10, 13], and phononic crystal approaches [14]. These approaches exhibit challenges in fabrication, extension to higher dimensions, and bandwidth, and thus the search for other techniques is ongoing [15].

In this work we have designed, fabricated and measured a labyrinthine metamaterial based on a previously described theoretical approach [16]. Both two-dimensional measurements of the interaction of an acoustic beam with a prism made of the labyrinthine metamaterial and extracted parameters from one-dimensional reflection-transmission measurements demonstrate that the structure possesses a broad frequency band of significantly negative refractive index.

Using a mechanism different from most of the above mentioned approaches which exploit local resonance to achieve unusual effective material parameters, a maze-like or labyrinthine structure has drawn attention as an unconventional acoustic metamaterial design [16]. In applied acoustics, folded structures have been used for the acoustic wave modulation, such as for modulating the phase of the backward wave and guiding it to superpose front radiated wave for enhancement [17]. Li and coworkers recently proposed a space-coiling metamaterial design [16] which we refer to as labyrinthine structure. Simulations confirmed the negative index and modestly dispersive nature of the structure and demonstrated its effectiveness as a building block for negative and zero index devices.

Building on that work [16], we kept the basic symmetry and space coiling characteristics of the theoretical design, while accounting for fabrication and measurement feasibility. We then fine-tuned the dimensions of our unit cell to render its negative index to zero index frequency range fall into the allowed frequency band of our measuring platform, while keeping the dimension small enough in terms of wavelength so that the structure can be regarded as effective medium. For the estimation of the dispersion characteristics of the labyrinthine structure, we retrieved parameters numerically employing a method presented in [18] and [20]. All numerical simulations were performed using the Acoustic-Solid Interaction module of COMSOL Multiphysics.

Macroscopically, the unit cell was designed to be at least 5 times smaller than a wavelength in air in the frequency range of 2,000 Hz to 3,000 Hz. As we will see shortly the effective wavelength inside a bulk material obtained by replicating the unit cell varies between approximately 3 unit cell diameters at 2,000 Hz and infinite close to 2,800 Hz where the refractive index drops to zero. This, together with experiments and simulations we are going to describe below, demonstrates that a material based on this unit cell is in the effective medium region and, thus, can function as a homogeneous material.

In a microscopic view, the labyrinthine structure forces the acoustic wave to travel in folded channels, enabling the path length of the wave to be multiplied. Band structure analysis [16] on the structure predicted the abundant existence of unusual refractive index values, including negative index, zero index and smaller-than-unity index. The resonating nature results from space-folding instead of local resonance inside the unit cell and brings two advantages: first, it creates an extraordinarily broad frequency range (more than 1,000 Hz for our design) of negative index, whereas for traditional locally resonating structure, the negative index can only be found in a narrow frequency band around the resonating peak; second, it circumvents the high absorption loss adhesively associated with a locally resonating peak. The dimensions of the unit cell and a photograph of a fabricated cell are shown in Fig. 1. The dispersive characteristics can be easily tailored by carefully designing the geometry of the cross section of the unit cell.

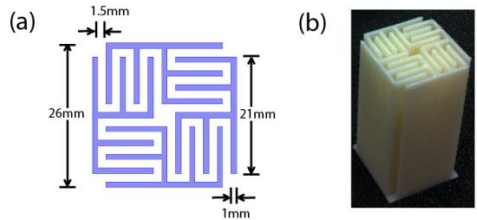

FIG. 1 (color online). Labyrinthine unit cell: (a) Designed dimensions of the unit cell. (b) Photograph of a fabricated unit cell

The labyrinthine unit cells were fabricated with thermoplastics via 3D printing. Firstly, we experimentally verified the negative refractive index of the labyrinthine unit cell by retrieving its effective material parameters inside a one-dimensional waveguide using a method similar to that presented in [19]. To perform the measurement, we built an 8 feet long rectangular waveguide made of rigid plastic. The waveguide dimensions were chosen such that one unit cell filled the entire transverse section of the waveguide. The cell was placed in the middle of the waveguide. Short (approximately 5 periods at the center frequency), modulated Gaussian pulses having center frequencies varying from 2,000 Hz to 3,000 Hz in steps of 50 Hz were sent through the unit cell, and the reflected and transmitted pulses were recorded by two microphones situated 25 inches before and, respectively, 9 inches after the unit cell. These distances were chosen such that the reflected and transmitted pulses could be easily isolated from the incident pulse and other subsequent reflections from the waveguide ends. These isolated transmitted and reflected pulses were Fourier transformed and calibrated with similar signals obtained when the cell was replaced by two calibration standards in order to obtain the complex reflection and transmission coefficients (or S-parameters). The two calibration standards were, first, a section of empty waveguide the same size as the unit cell. In this case the reflection coefficient is 0 and the transmission coefficient is $\exp(-j\omega d/v)$, where $\omega$ is the angular frequency, $d$ is the cell diameter, and $v$ is the speed of sound in air. The second standard is a perfect reflector in the form of a metal plate filling the entire cross-section of the waveguide, i.e. reflection coefficient of 1. The calibrated S-parameters were inverted using the method outlined in [18].

A comparison between the complex refractive indexes ($n = n'+in"$) obtained in numerical simulations and retrieved experimentally using the method outlined above is given in Fig. 2. In the negative index region, the measured refractive index changes towards zero as frequency increases, and both the real part and the imaginary part of the measured values agree well with the simulated values. Note that the imaginary part of the refractive index ($n"$) is too small to be reliably measured by our retrieval system, therefore the fluctuation of $n"$ above and below zero is consistent with the close-to-zero value obtained in simulation. The simulated and measured values of the refractive index show that the effective wavelength in a material generated by this unit cell varies between approximately 3 unit cell diameters at 2,000 Hz and very large values around 2,800 Hz where the refractive index approaches zero. This is the first indication that effective medium theory holds for materials based on this cell.

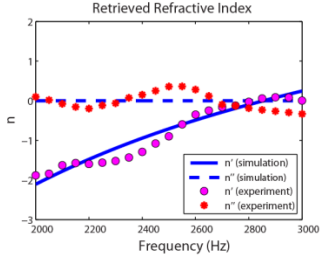

FIG. 2 (color online). Retrieved refractive index: (a) Retrieved index from experiment (pink circles for real part, red asterisks for imaginary part). (b) Retrieved index from simulation (solid blue line for real part, dashed blue line for imaginary part).

The one-dimensional parameters retrieved experimentally verify the theoretically predicted broad negative index band. In order to show that the labyrinthine unit cells are able to generate metamaterials characterized by well defined effective material parameters and usable in real devices, we performed a two-dimensional prism-based measurement to demonstrate negative refraction. A total of 55 identical unit cells were assembled to form the prism shown in Fig. 3(a). In our prism, we turned neighboring unit cells upside-down, so that the prism top and bottom would be more symmetric, and the gap between unit cells would be lessened. Note that, even though each flipped unit cell is now the mirror image of the unflipped one, the dispersion characteristics remain the same. The top and bottom surfaces of the prism are then sealed to minimize wave pressure leakage.

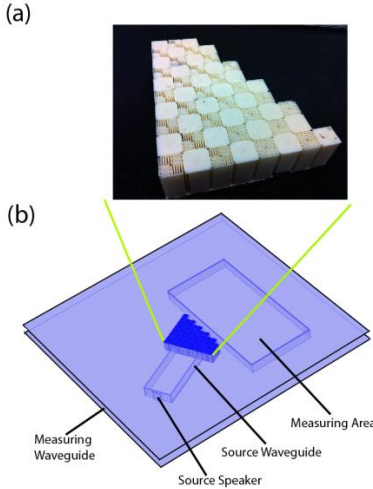

FIG. 3 (color online). Experimental setup: (a) Photograph of the assembled prism with 55 identical labyrinthine unit cells. (b) Two-dimensional measuring platform which was used to perform the field mapping.

The measurement was performed in our two-dimensional acoustic measuring platform [6,21] as shown in Fig. 3(b). The measuring waveguide is composed of two rigid plates separated by 2 inches. The acoustic waves propagate inside the waveguide in between the two plates. A source speaker was programmed to emit a short acoustic pulse with a Gaussian envelop in time and frequency, and was placed at one end of a rectangular smaller waveguide, which we call source waveguide, positioned inside the measuring waveguide. The source waveguide with its sound hard boundaries was used to send guided quasi-plane waves into the prism structure. The signal exiting from the prism was detected by a microphone step-sweeping over the measuring area [the rectangular region behind the prism in Fig. 3(b)]. At the data processing stage, the phase and amplitude at each measuring point were calculated and the field pattern was then plotted. Both the experimental setup and the method of mapping the acoustic field inside the waveguide are described in detail in [6] and [21].

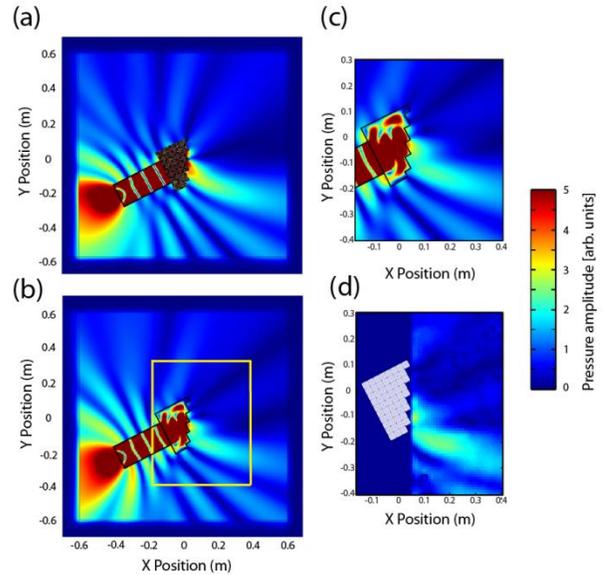

Fig. 4 (color online). Simulation and measurement showing negative refraction at the prism-air interface: (a) Simulation of prism made of the real labyrinthine structure at 2,100 Hz. (b) Simulation of the prism made of a with homogeneous material having the effective parameters retrieved from one-dimensional reflection/transmission measurements at 2,100 Hz. (c) Zoomed-in plot of the highlighted region in (b). (d) Measured field pattern at 2,100 Hz.

Two prisms were simulated and compared. The first one is made of the real labyrinthine structure, and the other one an identically shaped prism composed of a homogeneous material whose parameters were retrieved using the one-dimensional unit cell measurements described above (we will call the former 'real structure prism' and the latter 'effective medium prism' later on). The simulation results at 2100 Hz for these two cases are shown in Fig. 4(a) and Fig. 4(b), respectively. The close resemblance between these two field patterns further validates the effectiveness of treating our labyrinthine prism as a homogenous effective one. The measured field pattern at 2,100 Hz is shown in Fig. 4(d) and is compared with the simulation [Fig. 4(c), which shows the pattern obtained in the highlighted region of Fig. 4(b)]. The measurements clearly demonstrate that the prism bends the wave towards the same side as the incident wave with respect to the normal to the air/prism interface, implying negative refraction. The measured field pattern is in excellent agreement with the expected pattern obtained in simulations of the real structure prism and the effective medium prism.

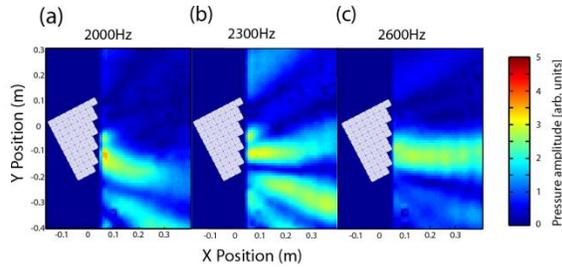

FIG. 5 (color online). Dispersion characteristics of the prism: Measured field pattern at (a) 2,000Hz , (b) 2,300Hz and (c) 2,600Hz respectively.

Examining the field patterns at frequencies ranging from 2,000 Hz to 2,700 Hz, we find a modestly dispersive negative index behavior of our prism. As the frequency increases, the exiting beam shifts its energy upwards towards the normal to the air/prism interface [see for example, the field pattern at 2,600 Hz shown in Fig. 5(c)], indicating an increase of the refractive index from negative values towards zero. At certain frequencies, we observed that the outgoing beam splits into multiple lobes [2,300 Hz shown in Fig. 5(b) is an example]. Simulations of the prism composed of continuous and homogeneous materials show the same behavior. This multi-beam phenomenon results mainly from the relatively small size (compared to the wavelength) of the prism.

To conclude, we have experimentally demonstrated that a properly designed labyrinthine acoustic metamaterial can exhibit a broad frequency range of negative refractive index and possesses attractive modest frequency dispersion. The geometry of the structure renders it relatively easy to tailor the effective properties and suitable for fabrication with rapid prototyping technology. The design principle can also be easily extended to higher dimensions to achieve three-dimension acoustic metamaterials. Therefore, we expect that this metamaterial design will facilitate the experimental realization of many theoretical proposals of acoustic metamaterial devices requiring negative or close-to-zero refractive index.

This work was supported by the Office of Naval Research through Grant N00014-12-1-0460.

*cummer@ee.duke.edu